\documentclass[12pt,preprint]{aastex}

\def\gta{\ifmmode {\mathbin{\lower 3pt\hbox   %> or of order
    {$\,\rlap{\raise 5pt\hbox{$\char'076$}}\mathchar"7218\,$}}}
    \else {${\mathbin{\lower 3pt\hbox
    {$\rlap{\raise 5pt\hbox{$\char'076$}}\mathchar"7218\,$}}}
    $}\fi}
\def\lta{\ifmmode {\,\mathbin{\lower 3pt\hbox   %< or of order
    {$\,\rlap{\raise 5pt\hbox{$\char'074$}}\mathchar"7218\,$}}}
    \else {${\mathbin{\lower 3pt\hbox
    {$\rlap{\raise 5pt\hbox{$\char'074$}}\mathchar"7218\,$}}}
    $}\fi}

\shorttitle {BURST RISE OSCILLATIONS FROM 4U 1636--536}
\shortauthors {Bhattacharyya and Strohmayer}

\begin{document}

\title {Evidence for Harmonic Content and Frequency Evolution of
Oscillations during the Rising Phase of X-ray Bursts from 4U
1636--536}

\author {Sudip Bhattacharyya\altaffilmark{1,2}, and Tod
E. Strohmayer\altaffilmark{2}}

\altaffiltext{1}{Department of Astronomy, University of Maryland at
College Park, College Park, MD 20742-2421}

\altaffiltext{2}{X-ray Astrophysics Lab,
Exploration of the Universe Division,
NASA's Goddard Space Flight Center,
Greenbelt, MD 20771; sudip@milkyway.gsfc.nasa.gov,
stroh@clarence.gsfc.nasa.gov}

\begin{abstract}

We report on a study of the evolution of burst oscillation properties
during the rising phase of X-ray bursts from 4U 1636--536 observed
with the proportional counter array (PCA) onboard the Rossi X-Ray
Timing Explorer (RXTE). We present evidence for significant harmonic
structure of burst oscillation pulses during the early rising phases
of bursts. This is the first such detection in burst rise
oscillations, and has interesting implications for constraining
neutron star structure parameters and the equation of state models of
matter at the core of a neutron star.  The detection of harmonic
content only during the initial portions of the burst rise appears
consistent with the theoretical expectation that with time the
thermonuclear burning region becomes larger, and hence the fundamental
and harmonic amplitudes both diminish.  We also find, for the first
time from this source, strong evidence of frequency increases during
burst rise.  The timing behavior of harmonic content, amplitude, and
frequency of burst rise oscillations may be important in understanding
the spreading of thermonuclear flames under the extreme physical
conditions on neutron star surfaces.

\end{abstract}


\section {Introduction} \label{sec: 1}

Millisecond period brightness oscillations, ``burst oscillations'',
during thermonuclear (type I) X-ray bursts from the surfaces of
neutron stars in low mass X-ray binary (LMXB) systems result from an
asymmetric brightness pattern on the rotating stellar surface
(Chakrabarty et al. 2003; Strohmayer, \& Bildsten 2003).  This timing
feature reveals the stellar spin period and can provide important
information about the other stellar parameters (radius, mass, etc.;
Miller, \& Lamb 1998; Nath, Strohmayer, \& Swank 2002; Muno, \"Ozel,
\& Chakrabarty 2002). Theoretical modelling of these oscillations has
the potential to constrain the equation of state (EOS) of the dense
matter at the core of a neutron star (see, for example Bhattacharyya
et al. 2005). This can be most effectively done if the burst
oscillation has some harmonic content, as the fitting of a pure
sinusoidal lightcurve does not put strong constraints on the stellar
parameters. Although burst oscillations have been discovered from more
than a dozen LMXBs, only one source (the accreting millisecond pulsar
XTE J1814-338) has shown significant harmonic content (during burst
decay; see Strohmayer et al. 2003; Watts, Strohmayer, \& Markwardt
2005). In this Letter we focus on oscillations during burst rise,
because at the beginning of the burst, the size of the burning region
is theoretically expected to be at its smallest, and the harmonic
content should theoretically be larger, and hence might be detected.

The study of burst oscillations during burst rise is important for
another reason.  At the onset of the burst, a small portion of the
stellar surface is ignited, and then the burning region spreads and
engulfs the whole stellar surface during the burst rise (Spitkovsky,
Levin, \& Ushomirsky 2002). This is natural, because simultaneous
ignition of the whole stellar surface would require very fine tuning.
Understanding this spreading is important, as it involves complex
nuclear physics and geophysical fluid dynamics (with significant
stellar spin), under conditions of extreme gravity, magnetic field and
radiation pressure. Other than the burst rise lightcurve, the time
evolution of three properties (frequency, amplitude, and harmonic
content) of burst oscillations are the most useful observational tools
to investigate this problem.  Studies to date have not found harmonic
content during burst rise.  Strohmayer, Zhang, \& Swank (1997) found
evidence for a decrease in amplitude during burst rise from an
analysis of RXTE data from the LMXB 4U 1728-34, and frequency increase
during the burst rise from the accreting millisecond pulsar SAX
J1808.4--3658 has also been reported (Chakrabarty et al. 2003).  In
this Letter, we report the finding of frequency evolution during the
rise of several bursts, as well as evidence for significant harmonic
content of burst oscillations at the beginning of burst rise.

\section {Data Analysis and results} \label{sec: 2}

More than a hundred bursts have so far been observed from the LMXB 4U
1636--536 by the proportional counter array (PCA) on board RXTE. By
analysing the archival data we find that 23 of them show at least
moderately strong oscillations during burst rise.  In order to search
sensitively for harmonic content, it is necessary to maximize the
fundamental power by taking any frequency evolution into account (see
Miller 1999; Strohmayer \& Markwardt 1999; Muno et
al. 2002). Therefore, first we explore frequency evolution during
burst rise using the following procedures: (1) we calculate dynamic
power spectra (Strohmayer \& Markwardt 1999) using a time duration
that is small enough to resolve the burst rise, but is still large
enough to accumulate significant signal power. The resulting dynamic
spectra (panel {\it a} of Fig. 1, and Fig. 2) provide an indication of
the frequency evolution behavior. (2) We carry out a phase timing
analysis (Muno et al. 2000) to confirm the indications in the dynamic
spectra.  We divide the burst rise time interval into several bins of
a fixed chosen length, and then assuming a frequency evolution model,
we calculate the average phase $(\psi_k)$ in each bin $(k)$. The
corresponding $\chi^2$ is calculated using the formula $\chi^2 =
\sum^M_{k=1} (\psi_k-\bar\psi_k)^2/\sigma^2_{\psi_k}$ (Strohmayer \&
Markwardt 2002), where $M$ is the number of bins, and $\sigma_{\psi_k}
= 1/\surd(Z_1^2)$ $(Z_1^2$ is the fundamental power in the
corresponding bin$)$.  We find the best fit parameter values for a
particular frequency evolution model by minimizing this $\chi^2$ and
we calculate the uncertainty in each parameter by increasing the
$\chi^2$ value by the appropriate amount (Press et al. 1992).  (3) To
confirm these results, we calculate the total fundamental power
$(Z_1^2)$ during the burst rise time interval using these best fit
frequency evolution model parameter values, and ensure that this power
is close to the maximum power obtained from any parameter values.

We found four bursts with significant frequency evolution during the
rise. These are listed in Table 1.  For burst A (panel {\it a},
Fig. 1), a constant frequency model gives a reduced $\chi^2 =
195.03/6$ and fundamental power $Z_1^2 = 57.41$. We also calculated
the phase residuals (see Strohmayer \& Markwardt 2002 for details) for
this model.  The large systematic deviations from the mean value
(panel {\it b}, Fig. 1), and the very high $\chi^2$ indicate that a
constant frequency model for burst A can be strongly rejected.  Next,
we add a linear term to the frequency model, and find that the
corresponding reduced $\chi^2 = 33.76/5$, and $Z_1^2 = 134.15$. This
is a better fit, but still not statistically acceptable. Therefore, we
include a nonlinear term to the frequency model. This model has a
reduced $\chi^2 = 2.00/4$, and $Z_1^2 = 170.42$, and the corresponding
phase residuals have small random deviations from the mean value
(panel {\it c}, Fig. 1). Hence, we conclude, that for burst A, the
data strongly indicate a nonlinear frequency increase (Table 1).  We
fit constant frequency models to the other bursts listed in Table
1. For bursts B, C, \& D, the reduced $\chi^2$ $(Z_1^2)$ values for
this model are 225.58/6 (86.91), 43.84/6 (86.16), \& 142.11/5 (20.42)
respectively. These are all poor fits, and hence the constant
frequency model for these bursts can also be strongly rejected.  Next,
we include a linear term in the frequency models for these bursts.
The corresponding reduced $\chi^2$ $(Z_1^2)$ values are 7.16/5
(200.35), 12.70/5 (173.18), \& 5.11/4 (89.75) respectively. These fits
are acceptable for bursts B \& D. For burst C, the high power value
and the good visual fit of this frequency model to the power contours
(panel {\it b}, Fig. 2) suggest that this model is on average correct,
and the high reduced $\chi^2$ value may be caused by fluctuation of
the frequency on short time scales. Therefore, for bursts B, C, \& D,
a linear frequency increase (Table 1 \& Fig. 2) is acceptable.  In
panel {\it d} of Fig. 1, we show the rms amplitude variation with time
during the rise of burst A. The fundamental amplitude shows an
initially fast and then a slower decrease.  Moreover, there is some
weak indication of a significant 1st harmonic amplitude during the
first $\sim 1/2$ second of this burst.

Using our best fitting frequency evolution models, we next searched
for harmonic content in the burst oscillations.  Individually, none of
the bursts shows strong harmonic power, therefore, we added the bursts
coherently to get more signal. For this purpose, we chose the bursts
with fundamental $Z^2$ power $> 30$ (for the constant frequency model)
during burst rise, and added them (nine in number; listed in Table 2)
coherently (i.e., we found the constant phase offset for each burst
which maximized the total fundamental power).  The total co-added,
phase-folded lightcurve does not give a significant harmonic power.  A
possible reason for this may be that during a significant portion of
the burst rise interval, the size of the burning region is large
enough that the harmonic amplitude is too small to be detected. To
address this possibility we consider five time intervals (starting at
the time of burst onset) of length; 1/4th, 1/3rd, half, 2/3rd, and all
of the rise time.  For each interval fraction, and for each burst, we
consider that fraction of the total rise time, fit it with our
frequency evolution model to get best fit parameter values, and use
these best fit values to calculate the phases. We do this for all the
nine bursts and then add them coherently (separately for each interval
fraction).  We find that the interval comprising 1/3rd of the rise
time gives the highest harmonic power $(17.52)$. The search at the
harmonic frequency in any independent subinterval is essentially a
single trial search (see Miller 1999), so we can estimate the
significance of this value using a $\chi^2$ distribution with 2
degrees of freedom. This gives a single-trial probability of
$1.57\times 10^{-4}$ to find a power as high by chance.  We searched 5
intervals, but they are not all independent, so the number of trials
is between 1 and 5.  Using 5 to get a conservative, lower bound on the
probability gives $7.85 \times 10^{-4}$, which is still a bit better
than a $3\sigma$ detection. As we increase the time interval, (that
is, use more of the rise), the harmonic power decreases gradually,
which is consistent with the expectation that the burning region
becomes larger, and hence the harmonic content diminishes. Due to the
consistency with this theoretical expectation, and the better than
$3\sigma$ harmonic power for 1/3rd of the rise interval, we conclude
that harmonic content in the burst rise oscillations has been
marginally detected.

We extracted the corresponding (i.e., for 1/3rd of the rise time
interval) phase-folded lightcurve from the data (after subtraction of
the persistent emission) of the nine bursts. We fit it with two
models: (1) a single sinusoid (the fundamental) around a constant
level, and (2) two sinusoids (fundamental and 1st harmonic) around a
constant level. The former one gives a reduced $\chi^2$ value
$26.26/13$, while this value for the latter model is $8.33/11$ (Table
2). These results support our finding that the addition of a 1st
harmonic provides a better description of the data.  Fig. 3 shows the
data, the best fit model (model 2 of Table 2), and all the components
of the model.

\section {Discussion}

In this Letter, we report two new observational results: (1) the first
detection of frequency evolution (increase) of burst rise oscillations
from 4U 1636--536, and (2) the first evidence for harmonic content of
burst rise oscillations (from any source). These effects may be a
direct result of thermonuclear flame propagation on the stellar
surface (Bhattacharyya \& Strohmayer 2005a; 2005b).  For example,
consider ignition of a burst off of the equator in the northern
hemisphere. Initially the small burning region (Spitkovsky et
al. 2002) can produce both a large fundamental and harmonic amplitude,
more or less consistent with the observations.  The frequency, when
first observed, is at its lowest value and then increases
monotonically. At least two effects can account for this behavior;
hydrostatic expansion---and subsequent spin-down---makes the burning
region slip westward (on a star rotating eastward; see Strohmayer,
Jahoda, Giles, \& Lee 1997; Cumming \& Bildsten 2001; Cumming et
al. 2002; Spitkovsky et al. 2002; Bhattacharyya \& Strohmayer 2005b),
and the southbound front will slip even further westward due to
conservation of angular momentum (Bhattacharyya \& Strohmayer 2005b).
Thus, the hot region initially has a net retrograde drift in the frame
of the neutron star, so the observed frequency is less than the spin
frequency.  As the front approaches the equator it spreads faster,
eventually forming a more or less symmetric equatorial belt
(Spitkovsky et al. 2002).  This can plausibly reduce the pulsation
amplitude in both the fundamental and harmonic, and the westward drift
slows because mass elements moving southward below the equator now
drift eastward, conserving angular momentum.  Thus, the frequency
increases from its initial (low) value.  Once the equatorial belt has
been ignited, it seems likely that residual asymmetry associated with
the initial, northbound burning front is responsible for the observed,
lower amplitude oscillations. Detailed calculations of the flux from
such a spreading burning front are required to explore this scenario
quantitatively, but the discussion above provides a plausible
qualitative understanding of many of the observed properties (more
detailed discussions are in Bhattacharyya \& Strohmayer 2005b).

As noted above, the harmonic content found at the beginning of the
bursts is consistent with the expected size of the burning region.  We
have computed theoretical models to show that the inferred harmonic
content is theoretically possible. For example, the harmonic to
fundamental amplitude ratio $(a_{\rm 2}/a_{\rm 1})$, and the relative
phase difference $(\epsilon_{\rm 1} - \epsilon_{\rm 2})$ of the
components can be reproduced reasonably well with a model assuming
emission from a circular hot spot (Bhattacharyya et al. 2005). Using a
dimensionless stellar radius-to-mass ratio $Rc^2/GM = 4.5$, stellar
mass $M = 1.5 M_\odot$, observer's inclination angle $i = 70^{\rm o}$,
$\theta$-position of the center of the circular burning region
$\theta_{\rm c} = 50^{\rm o}$, angular radius of the burning region
$\Delta\theta = 25^{\rm o}$, beaming parameter $n = 1.0$, and
blackbody temperature of the burning region $T_{\rm BB} = 2.0$~keV, we
can explain the relative strength and phasing of the fundamental and
harmonic components. Here, we have assumed a Schwarzschild spacetime,
and the beaming parameter $n$ gives a measure of the beaming in the
frame corotating with the star (Bhattacharyya et al. 2005).  These
model parameter values give $a_{\rm 2}/a_{\rm 1} = 0.24$ and
$\epsilon_{\rm 1} - \epsilon_{\rm 2} = 0.41$, while for the data
(Fig. 3), $a_{\rm 2}/a_{\rm 1} = 0.24\pm0.06$ and $\epsilon_{\rm 1} -
\epsilon_{\rm 2} = 0.36\pm0.02$. The constant level, required by the
observed lightcurve, can plausibly be supplied by a symmetric
belt-like component of the burning region (as mentioned earlier).
We note that waves in the surface layers of a neutron star (Heyl
2005; Lee \& Strohmayer 2005) are unlikely to produce a significant
harmonic content, and hence at least for the rising phases of these
bursts, a wave interpretation of oscillations seems disfavored.

Our results suggest that detailed studies of the onset of bursts
can, in principle, provide insight into the structure of neutron
stars, and how thermonuclear flames propagate on their surfaces.
Moreover, the studies of bursts with oscillations during both rise and
tail can determine whether these two oscillations are phase-connected,
and hence of the same origin.  Unfortunately, current studies are
observationally limited by the detected count rates. However, a way
out of this is to add a number of bursts to increase the signal to
noise ratio, as we have done in this Letter to detect the harmonic
content.

\acknowledgments

This work was supported in part by NASA Guest Investigator grants.

{}

\newpage

\begin{deluxetable}{cccccccc}
\tablecolumns{8}
\tablewidth{0pc}
\tablecaption{Frequency evolution model parameters\tablenotemark{a} (with 
1$\sigma$ error)
for four bursts.}
\tablehead{ObsId & Start date & Burst & $\nu_{\rm 0}$ & $\dot\nu$ & $\ddot\nu$ 
& $\chi^{\rm
2}$ & $Z^2_1$\tablenotemark{b}}
\startdata
60032-05-03-00 & 2002 Jan 12 & A & $577.70^{+0.24}_{-0.25}$ & 
$2.38^{+0.24}_{-0.25}$ &
$-0.42^{+0.07}_{-0.07}$ & $2.00$ & $170.42$ \\
\\
60032-05-03-00 & 2002 Jan 13 & B & $579.19^{+0.07}_{-0.07}$ & 
$0.81^{+0.06}_{-0.05}$ & --
& $7.16$ & $200.35$ \\
\\
60032-05-05-00 & 2002 Jan 14 & C & $580.62^{+0.12}_{-0.12}$ & 
$0.28^{+0.08}_{-0.08}$ & --
& $12.70$ & $173.18$ \\
\\
60032-05-10-00 & 2002 Jan 22 & D & $578.44^{+0.12}_{-0.13}$ & 
$1.21^{+0.11}_{-0.10}$ & -- &
$5.11$ & $89.75$ \\
\enddata
\tablenotetext{a}{Frequency evolution model: $\nu(t) = \nu_{\rm 0} + \dot\nu t 
+
\ddot\nu t^2$, where $\nu_{\rm 0} = \nu(0).$}
\tablenotetext{b}{Fundamental power during burst rise.}
\end{deluxetable}

\clearpage

\begin{deluxetable}{ccccccc}
\tablecolumns{7}
\tablewidth{0pc}
\tablecaption{Fitting of phase-folded lightcurve\tablenotemark{a}: best fit
parameter\tablenotemark{b} values (with 1$\sigma$ error).}
\tablehead{Model & $a_{\rm 0}$ & $a_{\rm 1}$ & $\epsilon_{\rm 1}$ & $a_{\rm 
2}$ &
$\epsilon_{\rm 2}$ & $\chi^2$/dof}
\startdata
1\tablenotemark{c} & $667.87\pm10.00$ & $241.42\pm14.23$ & $0.37\pm0.01$ & -- 
& -- &
$26.26/13$ \\
\\
2\tablenotemark{d} & $668.99\pm10.00$ & $245.60\pm14.26$ & $0.37\pm0.01$ & 
$59.49\pm14.05$ &
$0.01\pm0.02$ & $8.33/11$ \\
\enddata
\tablenotetext{a}{Combination (for first 1/3rd of the rise time interval)
of nine bursts from the following ObsIds (start date):
30053-02-02-02 (1998 Aug 19), 30053-02-02-00 (1998 Aug 20), 40028-01-02-00 
(1999 Feb 27),
40028-01-08-00 (1999 Jun 18), 50030-02-01-00 (2000 Nov 05), 60032-05-03-00 
(2002 Jan
12), 60032-05-03-00 (2002 Jan 13), 60032-05-05-00 (2002 Jan 14) and 
60032-05-13-00
(2002 Feb 05).}
\tablenotetext{b}{Model lightcurve intensity is $I = a_{\rm 0} + a_{\rm 1} 
\sin (2 \pi (\epsilon - \epsilon_{\rm 1})) + a_{\rm 2} \sin (4 \pi (\epsilon -
\epsilon_{\rm 2}))$, where $\epsilon$ is the phase variable.}
\tablenotetext{c}{Only fundamental.}
\tablenotetext{d}{Fundamental + 1st harmonic.}
\end{deluxetable}

\clearpage
\begin{figure}
\epsscale{.6}
\plotone{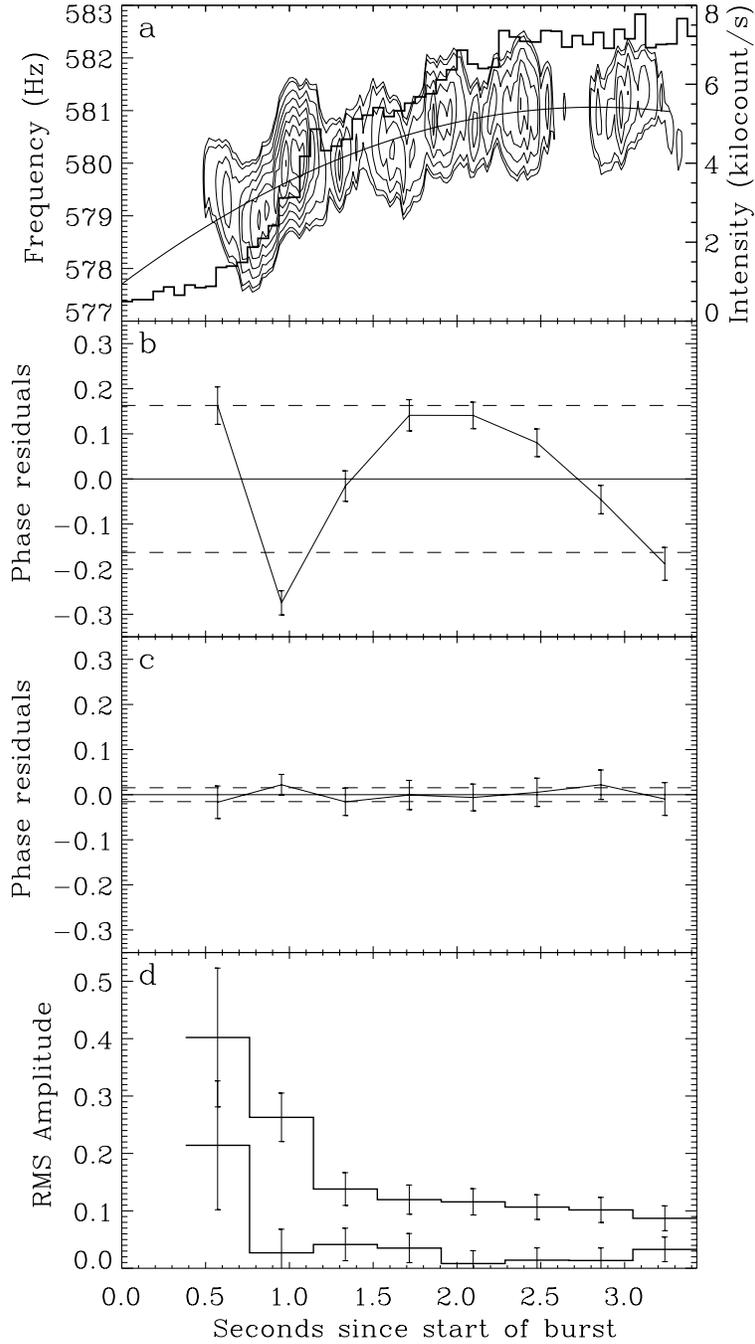}
\caption{Time evolution of different observed burst properties during
the rise of burst A (see Table 1) from 4U 1636--536. Panel {\it a}
gives the detected intensity (histogram), power contours (minimum and
maximum power values are $18$ and $55$) using dynamic power spectra
(for 0.4 s duration at 0.02 s intervals), and the best fit model from
Table 1.  Panel {\it b} gives the phase residuals (phase varies from 0
to 1) and rms deviation (broken horizontal lines) for constant
frequency $(580.72$~Hz) fitting. Panel {\it c} is same as panel {\it
b}, but for the best fit values of frequency evolution parameters
(Table 1). Panel {\it d} gives the rms amplitudes of brightness
oscillation (persistent emission subtracted). The upper histogram is
for the fundamental amplitude, and the lower histogram is for the 1st
harmonic amplitude. Here the horizontal lines give the binsize and the
vertical lines give $1\sigma$ error.}\end{figure}

\clearpage
\begin{figure}
\epsscale{.7}
\plotone{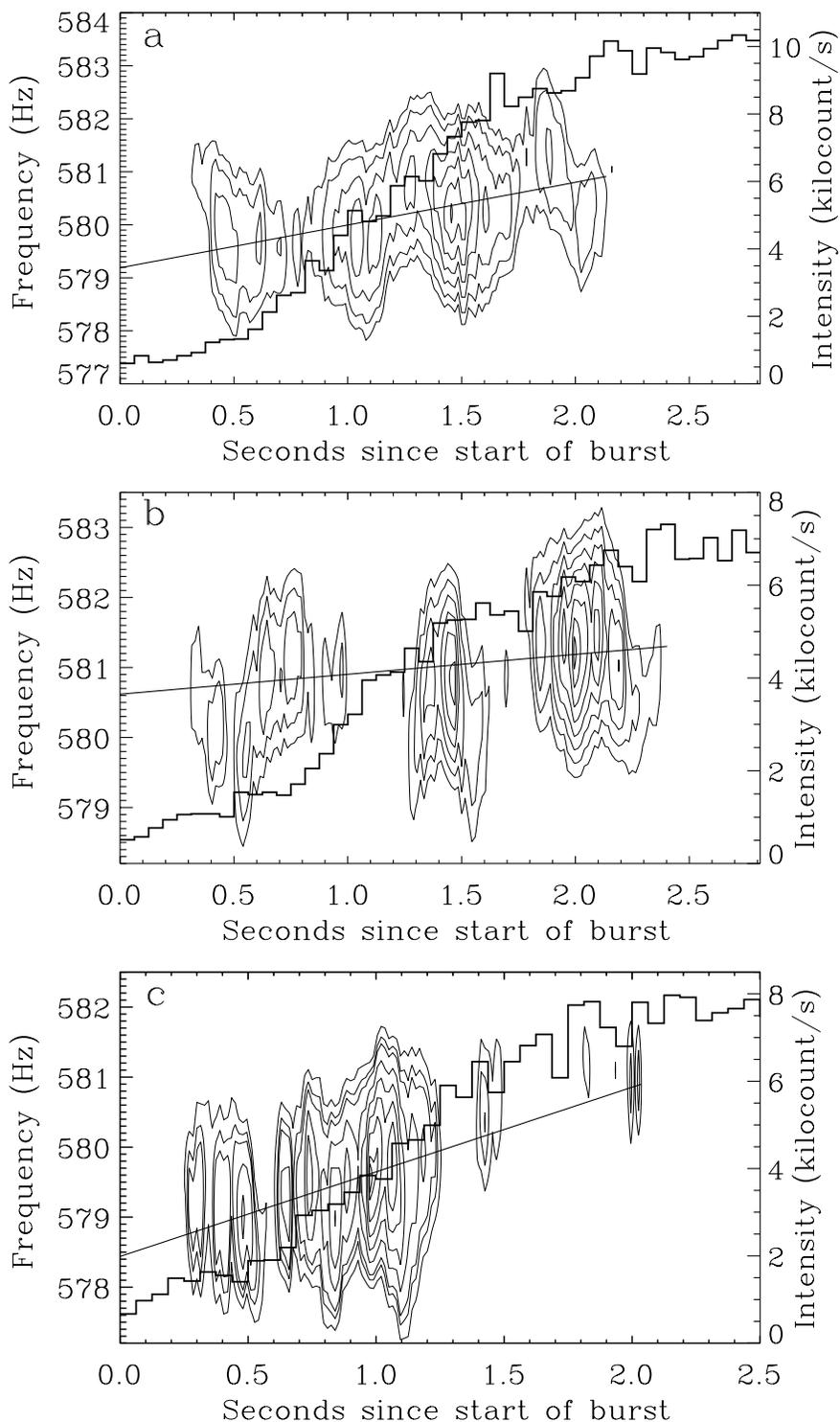}
\caption{Similar as panel {\it a} of Fig. 1 (panel {\it a} is for
burst B, panel {\it b} is for burst C, and panel {\it c} is for burst
D). For each of the panels, the dynamic power spectra are calculated
for 0.3 s duration at 0.015 s intervals. The minimum and maximum power
values of the contours are $(23,77)$, $(20,53)$, and $(15,46)$ for
panels {\it a}, {\it b}, and {\it c} respectively.}
\end{figure}

\clearpage
\begin{figure}
\epsscale{1.0}
\plotone{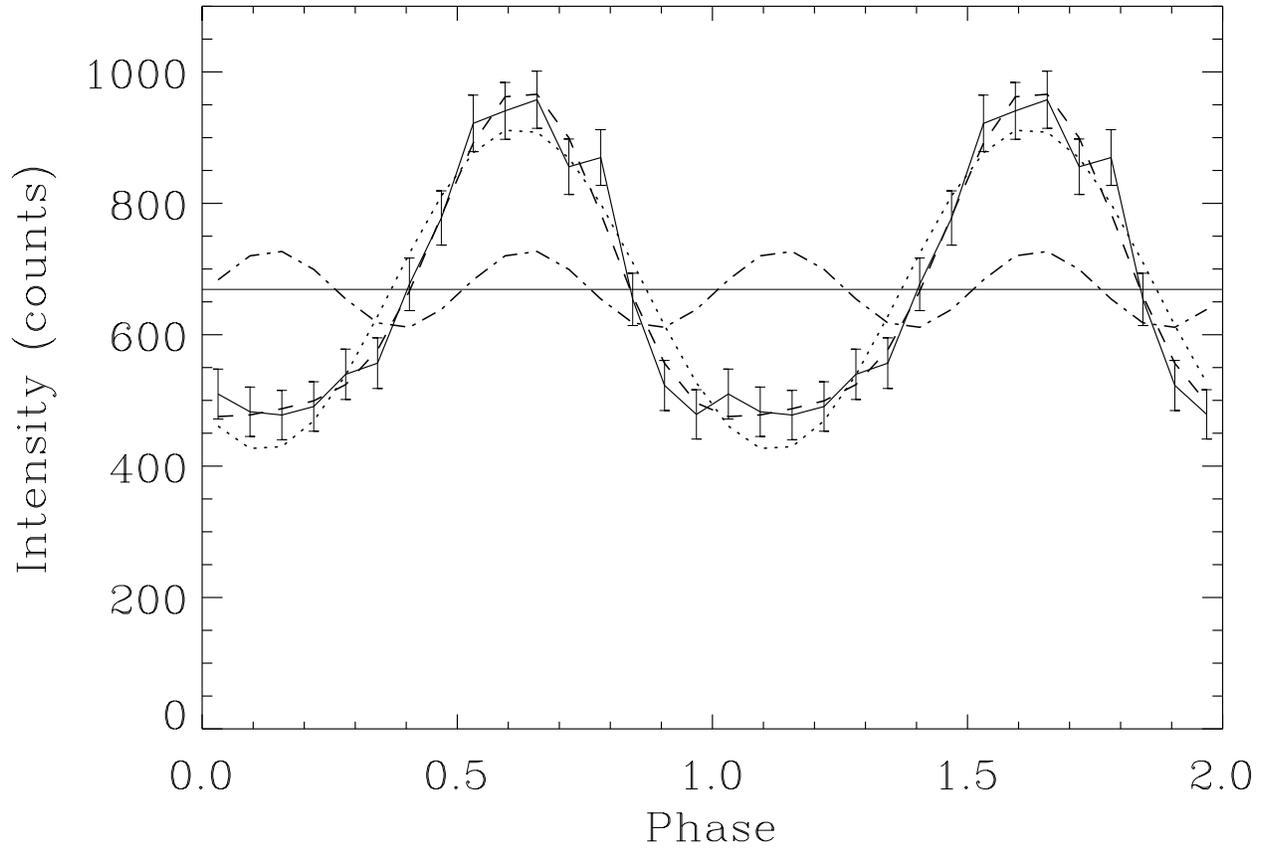}
\caption{Phase-folded lightcurve of burst oscillation during burst
rise from 4U 1636--536.  The solid curve shows the data (combined for
nine bursts for 1/3rd of the rise time interval; 
see Table 2), dashed curve is model 2 (see Table 2),
solid horizontal line is the constant level of the model, dotted curve
is the fundamental component of the model, and dash-dot curve is the
1st harmonic component of the model.}
\end{figure}

\end{document}